\documentclass[11pt,twoside]{article}

\usepackage{asp2014}

\aspSuppressVolSlug
\resetcounters

\begin{document}

\title{Identification of Artifacts and Interesting Celestial Objects
 in LAMOST Spectral Survey}

\author{Petr \v{S}koda,$^1$ Ksenia Shakurova,$^2$ Jakub Koza,$^2$ and Andrej Pali\v{c}ka$^2$}
\affil{$^1$Astronomical Institute of the Czech Academy of Sciences, Ond\v{r}ejov, Czech Republic; \email{skoda@sunstel.asu.cas.cz}}
\affil{$^2$Faculty of Information Technology, Czech Technical University, Prague, Czech Republic}

\paperauthor{Petr {\v{S}koda}}{skoda@sunstel.asu.cas.cz}{}{Astronomical Institute of the Czech Academy of Sciences}{Stellar department}{Ond{\v r}ejov}{}{25165}{Czech Republic}
\paperauthor{Ksenia Shakurova}{}{}{Faculty of Information Technology, Czech Technical University}{}{Prague 6}{}{16000}{Czech Republic}
\paperauthor{Jakub Koza}{}{}{Faculty of Information Technology, Czech Technical University}{}{Prague 6}{}{16000}{Czech Republic}
\paperauthor{Andrej Pali\v{c}ka}{}{}{Faculty of Information Technology, Czech Technical University}{}{Prague 6}{}{16000}{Czech Republic}

\begin{abstract}
The LAMOST DR1 survey contains about two million of spectra labelled by its
pipeline as stellar objects of common spectral classes. There is, however, a
lot of spectra corrupted in some way by both instrumental and processing
artifacts, which may mimic spectral properties of interesting celestial
objects, namely emission lines of Be stars and quasars. 

We have tested several clustering methods as well as outliers
analysis on a sample of one hundred thousand spectra using Spark scripts
running on Hadoop cluster consisting of twenty-four sixteen-core nodes. This
experiment was motivated by an attempt to find rare objects with interesting
spectra as outliers most dissimilar from all common spectra. 

The result of this time-consuming procedure is a list of several
hundred candidates where different artifacts are prominent, but also tens of
very interesting emission-line spectra requiring further detailed examination.
Many of them may be quasars or even blazars as well as yet unknown Be-stars.
It deserves mentioning that most of the work benefitted considerably from
technologies of Virtual Observatory.
\end{abstract}

\section{Finding Outliers with Unsupervised Machine Learning}

Machine learning is the field of informatics, closely related to the advanced
statistical inference, which tries to build models of data  by learning from
sample inputs and make predictions based on such learned models.  It is divided
mainly into supervised and unsupervised methods. 

Unsupervised learning (unlike supervised one requiring the labels assigned to
part of data) tries to identify similar patterns (typically different clusters
based on some similarity metrics) in data automatically without the human
intervention. The outliers are entities which cannot be assigned to any of such
cluster (so they represent the single member clusters).

In big spectral archives, where it is almost impossible to investigate every
spectrum visually, the yet unknown rare objects with strange features, or even
sources with yet undiscovered physical mechanism may be in principle found
using this method. In any case a lot of random instrumental artifacts will be
found as well as every one is unique and thus very rare. The artifacts caused
by systematic errors of the same nature, which repeats very often, may be
collected by clustering as well.

\section{LOF Method for Finding Outliers }

The Local Outlier Factor method (LOF) introduced by \citet{breunig2000lof} is based on an
idea to compare local density of~an~object to the local densities of its
neighbours.  The local density is estimated by the typical distance
$\varepsilon$ at which a point can be "reached" from its neighbors:

One of the key terms for LOF is the \textit{k-distance} and \textit{reachability distance} of \textit{k} nearest neighbours: for~any $k > 0 $ the \textit{k-distance} of object $p$ is the distance $d(p,o)$ between $p$ and an object $o \in D$ such that:
                \begin{itemize}
                        \item for at least $k$ objects $o^{\prime} \in D \setminus {p}$ it holds that $d(p,o^{\prime}) \le d(p,o)$;
                        \item for at most $k-1$ objects $o^{\prime} \in D \setminus {p}$ it holds that $d(p,o^{\prime}) < d(p,o)$. 
                \end{itemize}

\noindent It is the distance of the object $p$ to the $k$-th nearest neighbor, but set of the k nearest neighbor ($N_k(p)$) includes all objects at this distance (it can contain more than $k$ objects). Using \textit{k-distance} the \textit{reachability distance} can be defined as 

                \begin{equation}
                        \textit{reach-distance}_k(p,o) = \max ( \textit{k-distance}(o), d(p,o) )
                \end{equation}

\noindent The \textit{local reachability density} of object \textit{p} is defined as
                \begin{equation}
                        lrd_{k}(p) = 1 / \left( \frac{ \sum\limits_{o \in N_{k(p)}}^{} \textit{reach-distance}_{k}(p, o)} {|N_{k}(p)|} \right)
                \end{equation}
The local outlier factor of p is defined as
                \begin{equation}
                        LOF_{k}(p) = \frac { \sum\limits_{o \in N_{k(p)}} \frac{lrd_{k}(o)}{lrd_{k}(p)} }
                        {|N_{k}(p)|}
                \end{equation}
If the LOF is considerably larger than 1, the object is an outlier, if it is about 1, the object is comparable with others.

\section{ LAMOST Spectral Surveys}

The LAMOST telescope \citep{2012RAA....12.1197C} has been delivering  one of
currently largest mega-collections  of spectra (similar to Sloan Digital Sky
Survey). The sixteen LAMOST spectrographs are fed by 4000 fibres positioned by
micro-motors. Its publicly accessible Data Release 1
\citep[see][]{2015RAA....15.1095L} contains altogether \textbf{2\,204\,696}
spectra, with a spectral resolving power about 1800, covering the range 3690-9100
\AA. The LAMOST pipeline classified \textbf{1\,944\,329} of them as
stellar ones.

\section{Ond\v{r}ejov CCD700  Archive}
There is a lot of  objects in the Universe that  show specific  shapes of
some  important spectral lines, especially 
multiple emission lines caused by  circumstellar disk, such as Be and  B[e]
stars, cataclysmic variables or young stellar objects.

\noindent The unique source is the archive of spectra obtained with 700mm
camera of the coud\`e  spectrograph of the 2m Perek Telescope at Ond\v{r}ejov
observatory, a part of the Astronomical Institute of the Czech Academy of
Sciences. The archive (named CCD700) contains about twenty thousand spectra of
mainly Be stars and other emission-line objects exposed in spectral range
6250--6700~\AA{}  with spectral resolving power about 13000. 

\section{Input Spectra and Their Preprocessing}
The important part of data preparation before applying machine learning is the
data pre-processing. In our case the spectra have to be normalized to the
continuum (rectified), cut to the same wavelength range and re-binned into the
same grid of wavelength points.  This gives us the number of so called Feature
Vectors (FV). As we want to  compare the same algorithm on both Ond\v{r}ejov
CCD700  and LAMOST spectra, we have to cut  the LAMOST ones to the similar
wavelength range (about 6250--6750~\AA{}) as those from CCD700.  The result of
the preprocessing is the big CSV file with all spectral intensities
interpolated to the same wavelength grid.  This (big) CSV is uploaded on a
computing cluster running Spark.

\section{Massively Parallelized Processing Using Spark}
The Apache Spark is a set of libraries written in SCALA language, adapted for
calling from PYTHON, running on number of computing nodes in parallel. We have
used the academical cluster MetaCentrum consisting of twenty-four sixteen-core
nodes (the number of nodes assigned by the system is however unknown, dependent
on a availability and load of the cluster). 

The data were distributed across all nodes by HDFS filesystem of Hadoop
infrastructure. The special Spark-based version of LOF method was developed by
K.S. \citep{master_shakurova} for this task. The experiments were run on all, almost 20\,000
Ond\v{r}ejov CCD700 spectra and then on about 120\,000 spectra randomly
selected from those labelled as star in LAMOST DR1.

\section{Results}

Our experiments proved that the LOF method was able to find in the CCD700
archive all interesting cases of spectra like sharp emissions, asymmetric
double peak emission or even noisy late type stars spectra.  In the LAMOST data,
contaminated by a lot of spoiled spectra, it can identify those with random
instrumental artifacts, but also some spurious ones, which may represent interesting
objects deserving further investigation. An example on Fig.~\ref{lamost-M7}
shows the noisy spectrum classified by the LAMOST pipeline as late type M7
class star, however (after zooming) it clearly presents the combination of
absorption and emission profile in Oxygen OI lines seen typically
in Be stars (see Fig.~\ref{lamost-M7-zoom}).
\articlefigure[width=0.81\textwidth,height=0.3\textwidth]{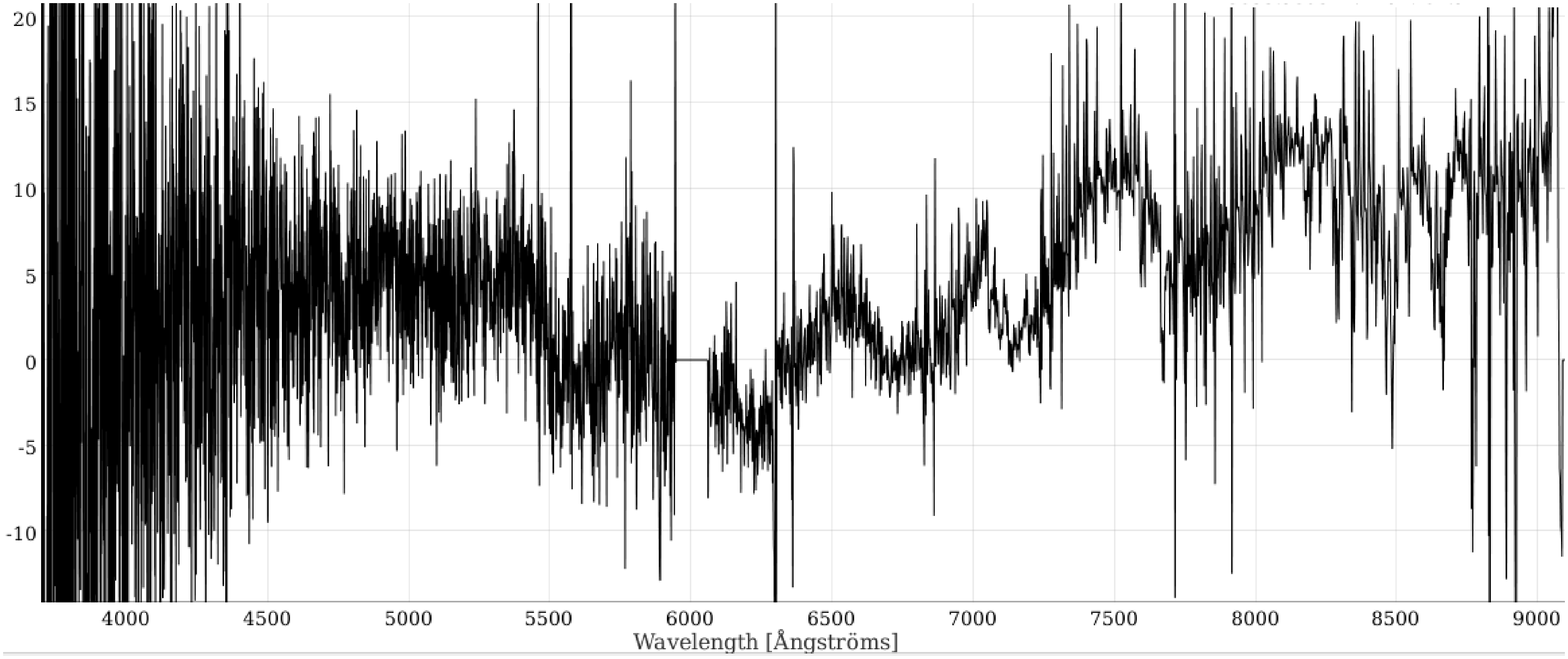}{lamost-M7}{LAMOST outlier classified as the M7 star}
\articlefigure[width=0.98\textwidth,height=0.33\textwidth]{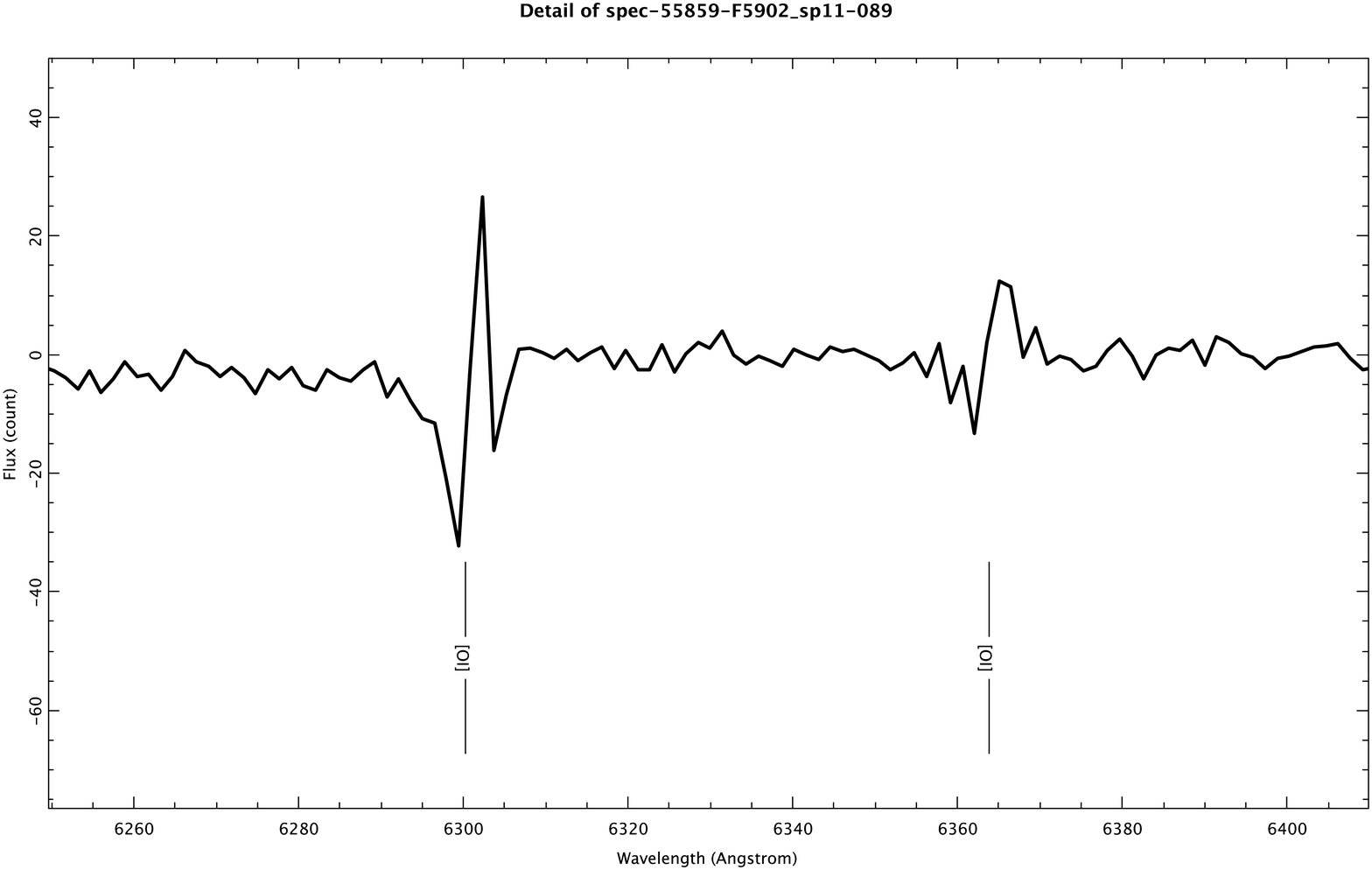}{lamost-M7-zoom}{Detailed view of emission in OI lines of the LAMOST outlier}

\section{Conclusions}
Big spectral archives are good source of  data suitable for machine learning of
interesting objects according to their characteristic spectral line shape. The
outlier finding methods as LOF may be successfully used for searching
instrumental artifacts but also the results need further detailed examination as
they may hide interesting  scientific objects. The application of the method
may  benefit considerably from massive parallelization using Spark on Hadoop
cluster.

\acknowledgements 
This work was supported by grant LD-15113 of Ministry of Education, Youth and
Sports of the Czech Republic.  This research is based on  spectra from
Ond\v{r}ejov 2m Perek telescope and public LAMOST DR1 survey.  Access to
computing and storage facilities owned by parties and projects contributing to
the National Grid Infrastructure MetaCentrum, provided under the programme
"Projects of Large Research, Development, and Innovations Infrastructures"
(CESNET LM2015042), is greatly appreciated.
\bibliography{P1-28.bib}  
\end{document}